\documentclass[epj,nopacs]{svjour}
\begin{document}
\title{Radiative Effects in Processes of
Diffractive Vector Meson Electroproduction}
\author{I. Akushevich
}                     
%
%
\institute{National Center of Particle and High Energy Physics,
Bogdanovich str. 153, 220040 Minsk, Belarus, e-mail: aku@hep.by}
\date{Received: 12 August 1998}
%
\abstract{
The electromagnetic radiative correction
to the cross section of the vector
meson electroproduction is calculated.
Explicit covariant
formulae for the observed cross section are obtained.
The dependence of the radiative correction on the experimental resolution
and on the inelasticity cut is discussed.
FORTRAN code DIFFRAD based on both
exact
(ultrarelativistic) and approximate sets of the formulae
for the radiative correction to the cross section is presented.
The detailed numerical analysis for kinematical conditions of the recent
experiments on the diffractive electroproduction of vector mesons is
given.
%
} 
\maketitle

\input{epsf.tex}

\section{Introduction}
\label{intro}
The measurement of the cross section of the exclusive vector meson
electroproduction can provide information on the had\-ronic
component of the photon and on nature of diffraction.
During several
years the diffractive production of the vector meson has been the
subject of the muonproduction \cite{EMC,NMC,E665} and
electroproduction
\cite{H1,ZEUS,HERMES} experiments.
Data analysis of these experiments is considerably affected by the QED
radiative effects.
At practice the radiative corrections (RC) to the processes of
electroproduction are taken into
account
using codes originally developed for the inclusive case (see
ref.\cite{Kurek}, for example).

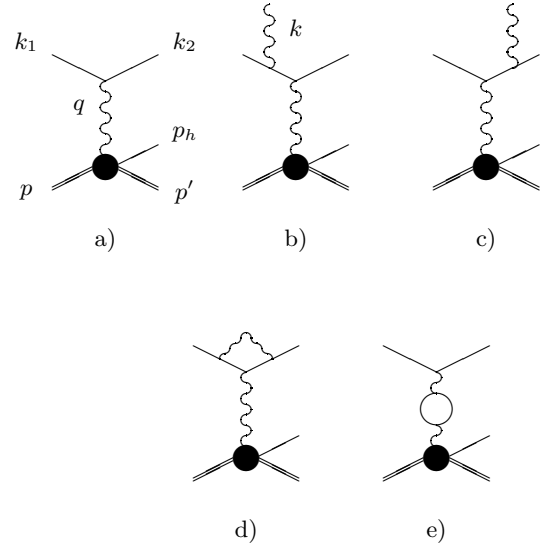
\begin{figure}[t]
\begin{tabular}{ccc}
\begin{picture}(60,100)
\put(30,60){\line(2,-1){20.}}
\put(50,50){\line(2,1){20.}}
\put(50,17.5){\circle*{10.}}
\multiput(50,28)(0,8){3}{\oval(4.0,4.0)[r]}
\multiput(50,24)(0,8){4}{\oval(4.0,4.0)[l]}
\put(30,10){\line(2,1){15.}}
\put(30,9){\line(2,1){15.}}
\put(55,18.5){\line(2,1){15.}}
\put(55,17.5){\line(2,-1){15.}}
\put(55,16.5){\line(2,-1){15.}}
\put(50,-10){\makebox(0,0){\small a)}}
\put(20,65){\makebox(0,0){\small $k_1$}}
\put(80,65){\makebox(0,0){\small $k_2$}}
\put(20,8){\makebox(0,0){\small $p$}}
\put(80,8){\makebox(0,0){\small $p'$}}
\put(80,30){\makebox(0,0){\small $p_h$}}
\put(40,40){\makebox(0,0){\small $q$}}
\end{picture}
&
\begin{picture}(60,100)
\multiput(40,57)(0,8){3}{\oval(4.0,4.0)[r]}
\multiput(40,61)(0,8){3}{\oval(4.0,4.0)[l]}
\put(30,60){\line(2,-1){20.}}
\put(50,50){\line(2,1){20.}}
\put(50,17.5){\circle*{10.}}
\multiput(50,28)(0,8){3}{\oval(4.0,4.0)[r]}
\multiput(50,24)(0,8){4}{\oval(4.0,4.0)[l]}
\put(30,10){\line(2,1){15.}}
\put(30,9){\line(2,1){15.}}
\put(55,18.5){\line(2,1){15.}}
\put(55,17.5){\line(2,-1){15.}}
\put(55,16.5){\line(2,-1){15.}}
\put(50,-10){\makebox(0,0){\small b)}}
\put(50,70){\makebox(0,0){\small $k$}}
\end{picture}
&
\begin{picture}(60,100)
\multiput(60,57)(0,8){3}{\oval(4.0,4.0)[r]}
\multiput(60,61)(0,8){3}{\oval(4.0,4.0)[l]}
\put(30,60){\line(2,-1){20.}}
\put(50,50){\line(2,1){20.}}
\put(50,17.5){\circle*{10.}}
\multiput(50,28)(0,8){3}{\oval(4.0,4.0)[r]}
\multiput(50,24)(0,8){4}{\oval(4.0,4.0)[l]}
\put(30,10){\line(2,1){15.}}
\put(30,9){\line(2,1){15.}}
\put(55,18.5){\line(2,1){15.}}
\put(55,17.5){\line(2,-1){15.}}
\put(55,16.5){\line(2,-1){15.}}
\put(50,-10){\makebox(0,0){\small c)}}
\end{picture}
\end{tabular}
\begin{center}
\begin{tabular}{rr}
\begin{picture}(60,100)
\put(30,60){\line(2,-1){20.}}
\put(50,50){\line(2,1){20.}}
\put(50,17.5){\circle*{10.}}
\multiput(50,28)(0,8){3}{\oval(4.0,4.0)[r]}
\multiput(50,24)(0,8){4}{\oval(4.0,4.0)[l]}
\multiput(42,55)(4,4){3}{\oval(4.0,4.0)[lt]}
\multiput(42,59)(4,4){2}{\oval(4.0,4.0)[br]}
\multiput(50,63)(4,-4){3}{\oval(4.0,4.0)[tr]}
\multiput(54,63)(4,-4){2}{\oval(4.0,4.0)[bl]}
\put(30,10){\line(2,1){15.}}
\put(30,9){\line(2,1){15.}}
\put(55,18.5){\line(2,1){15.}}
\put(55,17.5){\line(2,-1){15.}}
\put(55,16.5){\line(2,-1){15.}}
\put(50,-10){\makebox(0,0){\small d)}}
\end{picture}
&
\begin{picture}(60,100)
\put(30,60){\line(2,-1){20.}}
\put(50,50){\line(2,1){20.}}
\put(50,17.5){\circle*{10.}}
\multiput(50,48)(0,8){1}{\oval(4.0,4.0)[r]}
\multiput(50,44)(0,8){1}{\oval(4.0,4.0)[l]}
\multiput(50,28)(0,8){1}{\oval(4.0,4.0)[r]}
\multiput(50,24)(0,8){1}{\oval(4.0,4.0)[l]}
\put(50,36){\circle{12.}}
\put(30,10){\line(2,1){15.}}
\put(30,9){\line(2,1){15.}}
\put(55,18.5){\line(2,1){15.}}
\put(55,17.5){\line(2,-1){15.}}
\put(55,16.5){\line(2,-1){15.}}
\put(50,-10){\makebox(0,0){\small e)}}
\end{picture}
\end{tabular}
\end{center}
\vspace{0.5cm}
\caption{
\protect\it
Feynman diagrams contributing to the Born and the next order
cross sections. The letters denote the four-momenta of corresponding
particles.
}
\label{feyn}
\end{figure}

 The
purpose of this paper is to calculate the electromagnetic
correction to experimentally observed cross sections for the
kinematics of a fixed target and collider experiments directly.
The Feynman diagrams  necessary  to calculate
RC are presented on
Fig.\ref{feyn}.

In order to calculate exactly the QED RC to the cross section of
vector meson production, the method offered in ref.\cite{BSh} is used.
 By exact formulae is meant
 the expressions for the lowest order RC obtained
without
any approximations but ultrarelativistic: the lepton mass $m$ is
considered
to be small.
In  Section \ref{Exa} the kinematics of the radiative
and non-radiative processes and exact formulae for the
lowest order  RC
are obtained. In Section
\ref{Apr} the analytical results
are visualized by the construction
of the approximate formulae for cases interesting in practice.
The numerical results are given in Section \ref{Num}.
A brief discussion and  conclusions are
given in Section \ref{Dis}.

\section{Exact formulae for the lowest order correction}\label{Exa}

Seven kinematical variables are necessary to
describe the
radiative process of the diffractive vector meson production (Figure
\ref{feyn}b,c). Four of them are the same as for non-radiated case:
usual scaling variables $x$ and $y$,  the negative square of momentum
transfered from the virtual photon to the proton $t=(q-p_h)^2$ and
the angle $\phi_h$ between the scattering ($\vec{k_1}, \vec{k_2}$)
 and production ($\vec{q},\vec{p_h}$) planes in the laboratory
frame. The squared  virtual photon momentum $Q^2=-(k_1-k_2)^2$ and the
invariant mass of the initial proton and the virtual photon $W^2=(p+q)^2$
are
often used
instead of $x$ and  $y$.
The kinematics
of a real photon is described by three additional variables:   the
inelasticity $v={\Lambda}^2-M^2$ ($M$ is the proton  mass and
${\Lambda}^2=(p'+k)^2$ is the squared invariant mass
 of the system of unobserved particles),
$\tau=kq/kp$ and angle $\phi_k$
between planes
($\vec{k_1},\vec{k_2}$) and ($\vec{q},\vec{k}$).

We consider RC
to three and four dimensional cross sections
${\sigma}= d\sigma/dxdydtd\phi_h$ and
${\bar\sigma}= d\sigma/dxdydt$. They are related as
\begin{equation}
{\bar\sigma}=\int\limits_0^{2\pi} d\phi_h \; \sigma \; .
\label{eq0}
\end{equation}
 The four differential Born cross section
can be presented
 in the form
\begin{equation}
\sigma_0 =
\frac{\alpha}{4\pi^2xy}
\biggl(y^2\sigma_T+2\bigl(1-y-\frac{1}{4}y^2\gamma^2\bigr)
(\sigma_L+\sigma_T)\biggr),
\label{born}\end{equation}
where $\sigma_T$ and $\sigma_L$ are differential cross sections of
the photoproduction, $\gamma^2=Q^2/\nu^2$ and $\nu$ is the virtual photon
energy.

The differential cross section of the radiative process has
the following form:
\begin{equation}
\sigma_R \sim { |M_b+M_c|^2\over 4\;k_1p} dv {d^3k\over
2k_0}
\delta((\Lambda-k)^2-M^2),
\end{equation}
where $\Lambda=p+q-p_h$, $M_{b,c}$ are matrix elements of
the processes given in Figure \ref{feyn}b,c. In order to extract the
infrared divergence into a separate term
we following \cite{BSh}, use the identity
\begin{equation}
\sigma_R = \sigma_R - \sigma_{IR}+\sigma_{IR}
=\sigma_F+\sigma_{IR},
\label{eq133}
 \end{equation}
where
$\sigma_F$ is finite for $k\rightarrow 0$ and
 $\sigma_{IR}$ is the infrared divergent part
\begin{equation}
\sigma_{IR} = \frac{\alpha}{\pi}\delta_R^{IR}\sigma_0
= \frac{\alpha}{\pi}(\delta_S+\delta_H)\sigma_0 .
 \end{equation}
The quantities $\delta_S$ and $\delta_H$ appear after splitting the
integration region over $v$
by the infinitesimal parameter $\bar v$:
\begin{eqnarray}
\delta_S&=&\frac{1}{\pi}\int\limits_0^{\bar v}dv
\int\frac{d^{n-1}k}{(2\pi\mu)^{n-4}k_0}F_{IR}\delta((\Lambda-k)^2-M^2)
,\nonumber\\
\delta_H&=&\frac{1}{\pi}\int\limits_{\bar v}^{v_m}dv
\int\frac{d^3k}{k_0}F_{IR}\delta((\Lambda-k)^2-M^2),
\label{inte}
\end{eqnarray}
where $\mu$ is an arbitrary parameter of the mass dimension,
$v_m$ is a maximal inelasticity
and
\begin{equation}
F_{IR}= {m^2\over (2kk_1)^2}+{m^2\over (2kk_2)^2}-{Q^2+2m^2\over
(2kk_1)(2kk_2)}.
\end{equation}
The way to calculate the integrals
like (\ref{inte}) has been offered in ref.\cite{BSh} (see also review
\cite{Bardinrev}). In our case we have
\begin{eqnarray}
\delta_S&=&2\biggl(P_{IR}+\log\frac{\bar v}{\mu
M}\biggr)(l_m-1)+\log\frac{S'X'}{m^2M^2}+S_{\phi},
\nonumber\\
\delta_H&=&2(l_m-1)\log\frac{v_m}{\bar v},
\end{eqnarray}
where $l_m=\log(Q^2/m^2)$. The quantities $S'=2\Lambda
k_1=S-Q^2-V_1$ and
$X'=2\Lambda k_2=X+Q^2-V_2$ are calculated using
$V_{1,2}=2(a_{1,2}
+b\cos\phi_h)$, where
\begin{eqnarray}
a_1&=&\frac{1}{2\lambda_q}(Q^2S_pS_t-(SS_x+2M^2Q^2)t_q),
\nonumber\\
a_2&=&\frac{1}{2\lambda_q}(Q^2S_pS_t-(XS_x-2M^2Q^2)t_q),
\nonumber\\
b&=&\frac{1}{\lambda_q}(Q^2S_t^2-S_tS_xt_q-M^2t_q^2-
m_v^2\lambda_q)^{1/2}
\nonumber\\
&&\quad \times (SXQ^2-M^2Q^4-m^2\lambda_q)^{1/2}.
\end{eqnarray}
The invariants are defined as
\begin{eqnarray}
&&S=2k_1p, \; X=2k_2p=(1-y)S,\; Q^2=Sxy,
\nonumber\\
&&S_{p,x}=S\pm X, \; S_t=S_x+t, \; t_q=t+Q^2-m_v^2,
\nonumber\\&&\lambda_q=S_x^2+4M^2Q^2.
\end{eqnarray}

The infrared terms $P_{IR}$, parameters $\mu$ and $\bar v$ as well as
the squared logarithms containing the mass singularity $l_m^2$
are completely
canceled in the sum of $\delta_R^{IR}$
 with $\delta_V$ coming from a contribution of the vertex
function (Figure \ref{feyn}d):
\begin{equation}
\delta_V=-2(P_{IR}+\log\frac{m}{\mu})(l_m-1)-\frac{1}{2}l_m^2+\frac{3}{2}l_m
-2+\frac{\pi^2}{6}.
\end{equation}
For this sum we have
\begin{equation}
\delta_S+\delta_H+\delta_V=\delta_{inf}+\delta_{VR},
\end{equation}
where
\begin{eqnarray}        \label{deltas}
\delta_{VR} &=&\frac{\alpha}{\pi}
\biggl(\frac{3}{2}l_m-2-\frac{1}{2}\log^2\frac {X'}{S'}
+{\rm Li}_2\biggl[1-\frac{Q^2M^2}{S'X'}\biggr]
-\frac{\pi^2}6\biggr),
\nonumber\\[0.2cm]
\delta _{inf}&=& \frac{\alpha}{\pi}
(l_m-1)\log\frac{v_m^2}{S'X'}.
\end{eqnarray}
Here the ultrarelativistic expression for $S_{\phi}$ calculated in
\cite{Sh} was used. The
higher order corrections can be partially taken into account using a
special procedure of exponentiation of the multiple soft photon radiation.
There is an uncertainty, what part of $\delta_{VR}$ has to be
exponentiated. Within the considered approach \cite{Sh}
($1+\delta_{inf}$) is replaced by $\exp \delta_{inf}$.

For the observed cross section of the vector meson electroproduction
we obtain
\begin {equation}
\sigma _{obs} = \sigma _0 e^{\delta_{inf}}
(1 + \delta_{VR}+\delta_{vac})+
\sigma_{F}.
\label{eq134}
\end {equation}

The correction
    $\delta_{vac}$  comes from the
effects of vacuum polarization by leptons and hadrons (Figure
\ref{feyn}e). The  explicit
QED formulae for the first one can be found in ref.\cite{ASh}.
 The hadronic contribution
is given by a fit coming
from the data on
$e^+e^- \rightarrow hadrons $ \cite{delvac}.

The contribution of the infrared finite part
 can be written
in terms of POLRAD
2.0 notation \cite{ASh,POLRAD20}:
\begin{eqnarray}\label{020}
\sigma _F &=& -{\alpha^2 y\over 16 \pi^3}
 \int\limits^{2\pi }_{0}d\phi_k
 \int\limits^{\tau_{max}}_{\tau_{min}}d\tau
\sum_{i=1}^2\sum_{j=1}^3 \theta_{ij}
\times
\nonumber\\&& \qquad \qquad
\times\int\limits^{v_{m}}_{0}
{dv\over f}
R^{j-2}
\biggl[
{{\cal F}_i \over {\tilde Q}^4 }
-\delta_j{{\cal F}_i^0 \over Q^4 }
\biggr]
,
\end{eqnarray}
where $R=v/f$, $f=1+\tau-\mu$ and
$2M\tau_{max,min}=S_x\pm \sqrt{\lambda_q}$; $\delta_j=1$ for
$j=1$ and  $\delta_j=0$ otherwise.

The quantities $\theta_{ij}$ depend only on the kinematical invariants
and the integration variables $\tau$ and $\phi_k$
\begin{eqnarray}
\theta_{11}&=& 4 Q^2  F^0_{IR}
,\nonumber\\
\theta_{12}&=& 4 \tau  F^0_{IR}
,\nonumber\\
\theta_{13}&=&  - 2 (2F + F_{d} \tau^2)
,\nonumber\\
\theta_{21}&=& 4 (SX-M^2Q^2) F^0_{IR}
,\\
\theta_{22}&=&  -  F_{d} S_p^2 \tau +  F_{1+} S_p S_x
+ 2 F_{2-} S_p
\nonumber\\&& - 4 F^0_{IR} M^2 \tau + 2 F^0_{IR} S_x
,\nonumber\\
\theta_{23}&=& 4  F M^2 + 2 F_{d} M^2 \tau^2 - F_{d}S_x \tau -
 F_{1+} S_p,\nonumber
\end{eqnarray}
where
\begin{equation}
F_{d}=\frac{F}{z_1z_2},
F_{1+}=\frac{F}{z_1}+\frac{F}{z_2},
F_{2\pm}=F\biggl(\frac{m^2}{z_2^2}\pm\frac{m^2}{z_1^2}\biggr),
\end{equation}
 $F=1/(2\pi\sqrt{\lambda_Q})$ and $F^0_{IR}=F_{2+}-Q^2F_d=
F_{IR}R^2$.
The quantities
$\mu=kp_h/kp=2(a_k+b_k\cos(\phi_h-\phi_k))$
and $z_{1,2}=kk_{1,2}/kp=2(a_{1,2}^z-b^z\cos\phi_k)$ include
the dependence on angles:
\begin{eqnarray}
a_1^z&=&\frac{1}{2\lambda_q}(Q^2S_p+\tau (SS_x+2M^2Q^2)),
\nonumber\\
a_2^z&=&\frac{1}{2\lambda_q}(Q^2S_p+\tau (XS_x-2M^2Q^2)),
\nonumber\\
a_k&=&\frac{1}{2\lambda_q}((2Q^2+\tau S_x)S_m-(S_x-2\tau M^2)t_q),
\nonumber\\
b^z&=&\frac{1}{\lambda_q}(SXQ^2-M^2Q^4-m^2\lambda_q)^{1/2}
\nonumber\\
&&\quad \times (Q^2+\tau S_x-\tau^2M^2)^{1/2},
\nonumber\\
b_k&=&\frac{1}{\lambda_q}(Q^2S_m^2-S_mS_xt_q-M^2t_q^2-m_v^2\lambda_q)^{1/2}
\nonumber\\
&&\quad \times (Q^2+\tau S_x-\tau^2M^2)^{1/2}.
\end{eqnarray}
Here $S_m=S_t-v$.

The dependence on the photoproduction cross sections is included in ${\cal
F}_i$:
\begin{equation}
\begin{array}{ll}
\displaystyle
{\cal F}_1=(S_x-R){\sigma}_{T}^R, &
\displaystyle
\;\;{\cal F}_2={2{\tilde Q}^2\over
S_x-R}({\sigma}_{T}^R+{\sigma}_{L}^R),
\\[0.3cm]
\displaystyle
{\cal F}_1^0=S_x\sigma_T, &
\displaystyle
\;\;{\cal F}_2=2x(\sigma_T+\sigma_L).
\end{array}
\end{equation}
The quantities ${\sigma}_{T,L} $ have to be calculated for Born
kinematics, but ${\sigma}_{T,L}^R$ is calculated in terms of
so called true kinematics. It means that they have to be calculated
for the tilde variables
\begin{eqnarray}\label{008}
 {\tilde Q^2}  &  = & Q^2+R\tau,    \nonumber\\
 {\tilde W^2} &  = & W^2-R(1+\tau), \nonumber\\
 {\tilde t}  &  = & t+R(\tau-\mu)
\end{eqnarray}
instead of usual $Q^2$, $W^2$ and $t$.

\begin{figure}[h]
\unitlength 1mm
\begin{picture}(80,80)
\put(69,6){\makebox(0,0){$-t$,GeV$^2$}}
\put(31,61){\makebox(0,0){\scriptsize $x$=0.05, $Q^2$=1 GeV$^2$}}
\put(31,53){\makebox(0,0){\scriptsize $x$=0.10, $Q^2$=2 GeV$^2$}}
\put(31,45){\makebox(0,0){\scriptsize $x$=0.15, $Q^2$=3 GeV$^2$}}
\put(5,0){
\epsfxsize=8cm
\epsfysize=8cm
\epsfbox{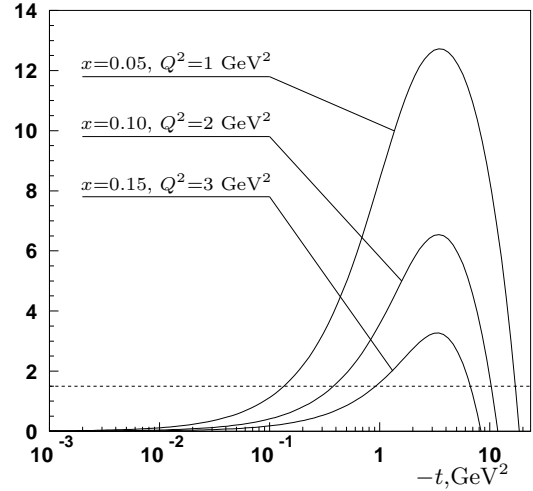}
}
\end{picture}
\caption{Maximal inelasticity within kinematical conditions of HERMES.
Dashed line gives a possible cut.}
\label{vmax}
\end{figure}

The important point is  the dependence of the results on the maximal
inelasticity
$v_{m}$. The
inelasticity is calculated in terms of the measured momenta, so it is
possible
to make a cut on the maximal value of this quantity.  If this cut is not
applied the maximal inelasticity is defined by kinematics only. Below we
give the formulae for
$ v_{m}$ in terms of the kinematical invariants
\begin{eqnarray}\label{vmkin}
4Q^2v_{m}&=&\Biggl(\sqrt{\lambda_q}-\sqrt{t_q^2+4m_v^2Q^2}\Biggr)^2-
\nonumber\\
&&
 \qquad
-(S_x-2Q^2+t_q)^2-4M^2Q^2
\end{eqnarray}
 and in terms of kinematical limits on $t$
\begin{equation}
   v_{m}=\frac{1}{C}{(t_{max}-t)(t-t_{min})},
\end{equation}
where $C$ behaves for small $t$ as
\begin{equation}
C={Q^2+m_v^2\over 2W^2}
\Bigl(S_x+\sqrt{\lambda_q}\Bigr)+O(t).
\end{equation}
The maximal inelasticity given by kinematics are plotted in Figure
\ref{vmax}.

\section{The approximation}
\label{Apr}

The structure of the exact formulae for the cross section of the
hard bremsstrahlung $\sigma_{F}$
 obtained in the last section  is not
simple.
Here we analyze them under approximation.  The third integration
variable $v$ is not completely photonic one.
It is calculated from the measured
momenta of the final lepton and the vector meson.
For non-radiated events $ v\equiv 0$. At
practice, however the distribution over
$v$ has a Gaussian form due to the
finite resolution. Radiative effects also affect the
distribution
(see sect. \ref{Num3}).
 So
 a cut on the invariant mass of the unobserved system rejects events with
hard radiated photon and  helps to reduce the total RC. Below we
consider
 the approximation based on assumption that $v_m$ is
relatively small or kinematical cut on $v$ is used.
 Besides, for the diffractive scattering
the squared  momentum $t$ transfered to the hadronic system is
always small, so
we construct the approximate formulae for RC to the cross section
$d\sigma/dxdydt$ with an additional assumption
\begin{equation}\label{appro}
M^2\ll S; \; t \ll Q^2.
\end{equation}
 It is clear that
this condition is normal for the collider experiments but it is
often not
so bad for
fixed target experiments as well. Due to the smallness of $v_m$ we can
keep only
leading term in the expansion of integrand (\ref{020}) over $v$.
First non-vanishing term contents the
first derivative of the integrand. The
main
contribution to this derivative comes from $\exp(b_vt)$ ($b_v$ is
a slope parameter). It is not difficult to
keep all the contributions to the derivative but for simplicity we
restrict our
calculation to this contribution. In this case
\begin{equation}\label{521}
 \frac{1}{v}\biggl[{Q^4\over {\tilde Q}^4}{{\cal F}_i
 \over{\cal F}_i^0}-1\biggr]
\approx b_v {\tau -\mu \over 1+\tau -\mu}
\end{equation}
and it is possible to integrate over $\phi_h$,
$\phi_k$ and
$\tau$
explicitly:
\begin{eqnarray}\label{522}
\frac{1}{2\pi}\int\limits_0^{2\pi}d\phi_h {\tau -\mu \over 1+\tau
-\mu}
&
\approx&
\frac{Q^2+m_v^2}{S_x-Q^2-m_v^2},
\nonumber\\
 \int\limits^{2\pi }_{0}d\phi_k
 \int\limits^{\tau_{max}}_{\tau_{min}}d\tau
 F_{IR}^0 &=& -2\bigl( l_m-1\bigr).
\end{eqnarray}

Final result for $\bar\sigma_F$ (see eq.(\ref{eq0})) is simple
\begin{equation}
{\bar \sigma_{F}}=
\frac{2\alpha b_v v_m}{\pi}(l_m-1)
\frac{Q^2+m_v^2}{S_x-Q^2-m_v^2}
{\bar \sigma_0}.
\label{051}
\end{equation}

Apart from a simple analytical form
the obtained formula has one more advantage. The
correction
depends only on the kinematical variables but not on the dynamical
characteristics
of the
interaction like  $\sigma_T$ and $\sigma_L$. It allows to eliminate
a possible
systematical error coming from the choice of the model. Such systematics
could be large because we have to know the differential cross sections
$\sigma_T$ and $\sigma_L$ in the wide region of varying of four variables, but
there are neither  enough experimental data for that nor satisfactory
models. Using (\ref{051}) we have systematics coming from the using of the
approximate
formulae instead of exact ones. But we are able to control it by comparing
the values calculated by (\ref{020}) and (\ref{051}) using any
model.

Moreover there
is a possibility to
 provide the procedure 'event
by event' reweighting each event with the RC factor. The comparison of
the results for the cross sections with and without this reweighting gives
the
correction.

\section{Numerical analysis and code DIFFRAD} \label{Num}

In this section we present the FORTRAN code DIFFRAD created on the
basis of the exact formulae performed in Section \ref{Exa}. The
program calculates the lowest order RC to the diffractive vector meson
electroproduction. The higher order effects are approximated by the
procedure of exponentiation. The formulae  for the cross section are given
in a covariant form, so the code can be run both for the fixed target
experiments and for the experiments at collider.
The   model
for $\sigma_{L,T}$ presented originally in ref.\cite{Ryskin} and
developed in ref.\cite{AR} is used as an input.

Below we give numerical results for RC factor
\begin{equation}
\eta={\bar\sigma_{obs}\over\bar\sigma_0}
={\int\limits_0^{2\pi}d\phi_h\sigma_{obs}\over
\int\limits_0^{2\pi}d\phi_h\sigma_0}
\end{equation}
and $t$- and $v$-distributions
obtained
within the kinematical regions of recent experiments on the
electro- and muonproduction of vector mesons.

\subsection{RC to the cross section}
\label{Num1}

\begin{figure}[b]
\centering
\unitlength 1mm
\parbox{.4\textwidth}{\centering
\begin{picture}(80,80)
\put(65,11){\makebox(0,0){$Q^2$,GeV$^2$}}
\put(20,75){\makebox(0,0){ $\eta$}}
\put(0,5){
\epsfxsize=8cm
\epsfysize=8cm
\epsfbox{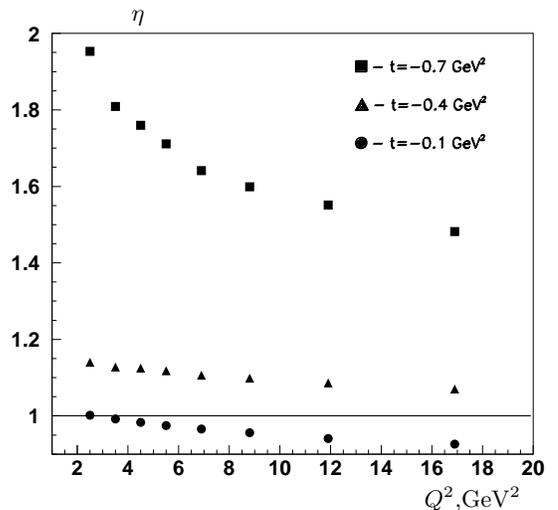}
}
\end{picture}
\vspace{-1.5cm}
\caption{RC factor in kinematical conditions of EMC/NMC
($\mu p \rightarrow \mu p \rho$).}
\label{f1}
}
\end{figure}

       On Figure \ref{f1} one can see $Q^2$-and $t$-dependences of
RC factor $\eta$
 in the kinematical region of
experiments of EMC and NMC. There is no cut on
inelasticity used. For high values of $t$ the RC  can reach
factor two. The reason of so large effect is smallness of the Born
photoproduction cross sections $\sigma_{L,T}$. They fall as
$exp(b_vt)$. In the observed cross section this factor is under
integral and there is a contribution from the region on $t$ where
$\sigma_{L,T}$
are not so small. As a result $\sigma_{obs}$ falls with the increasing
$-t$ but not so fast.

\begin{figure}[t]
\unitlength 0.5mm
\begin{picture}(165,165)
\put(55,160){\makebox(0,0){\small  $W^2=100$GeV$^2$}}
\put(130,160){\makebox(0,0){\small $W^2=200$GeV$^2$}}
\put(55,77){\makebox(0,0){\small   $W^2=400$GeV$^2$}}
\put(130,77){\makebox(0,0){\small  $W^2=600$GeV$^2$}}
\put(145,88){\makebox(0,0){\scriptsize $Q^2$,GeV$^2$}}
\put(65,88){\makebox(0,0){\scriptsize  $Q^2$,GeV$^2$}}
\put(145,16){\makebox(0,0){\scriptsize $Q^2$,GeV$^2$}}
\put(65,16){\makebox(0,0){\scriptsize  $Q^2$,GeV$^2$}}
\put(0,0){
\epsfxsize=9cm
\epsfysize=9cm
\epsfbox{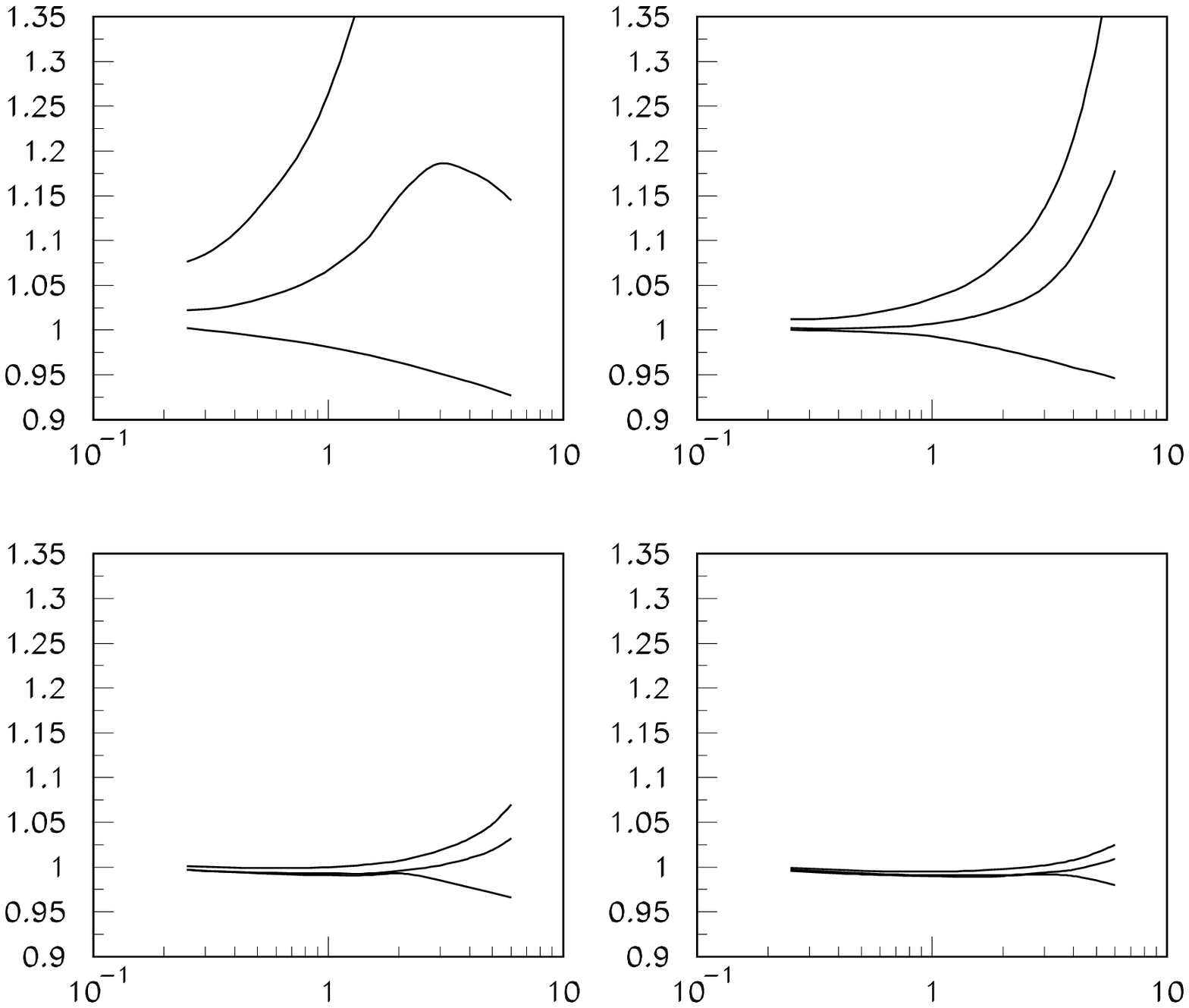}
}
\end{picture}
\vspace{-0.8cm}
\caption{RC factor in kinematical conditions of E665
($\mu p \rightarrow \mu p \rho$).
Curves from top to
bottom correspond to $t=$-0.9, -0.5, -0.1 GeV$^2$.}
\label{f2}
\end{figure}

Figure \ref{f2} gives the results for $\eta$ within kinematics of the
experiment E665 with a cut on the inelasticity. Notice it can be done
because of the rather  good
resolution over $v$  (standard deviation $\sigma $ of its
distribution is
more smaller than $v_{m}$ given by kinematics). The usage of this
cut leads to a different behavior of $\eta$ as a function of $Q^2$.
The different plots on this figure give $W^2$-dependence of $\eta$.

\begin{figure}[t]
\unitlength 0.5mm
\begin{picture}(165,165)
\put(60,150){\makebox(0,0){\scriptsize  $W^2=1000$GeV$^2$}}
\put(140,150){\makebox(0,0){\scriptsize $W^2=4000$GeV$^2$}}
\put(60,77){\makebox(0,0){\scriptsize   $W^2=8000$GeV$^2$}}
\put(140,77){\makebox(0,0){\scriptsize  $W^2=12000$GeV$^2$}}
\put(155,90){\makebox(0,0){\scriptsize  $Q^2$,GeV$^2$}}
\put(70,90){\makebox(0,0){\scriptsize   $Q^2$,GeV$^2$}}
\put(155,16){\makebox(0,0){\scriptsize  $Q^2$,GeV$^2$}}
\put(70,16){\makebox(0,0){\scriptsize   $Q^2$,GeV$^2$}}
\put(0,0){
\epsfxsize=9cm
\epsfysize=9cm
\epsfbox{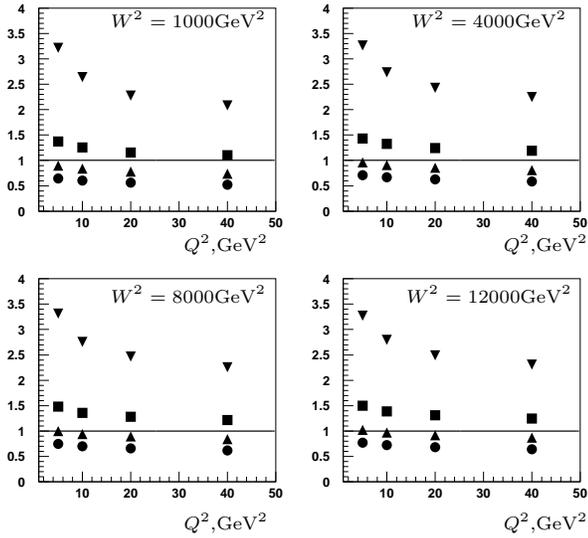}
}
\end{picture}
\vspace{-0.8cm}
\caption{RC factor in kinematical conditions of collider
experiments at HERA
($e p \rightarrow e p \rho$).
 Symbols from top to bottom correspond to
$t=$-0.7, -0.5, -0.3, -0.1 GeV$^2$}
\label{f3}
\end{figure}

\begin{figure}[t]
\centering
\unitlength 1mm
\parbox{.4\textwidth}{\centering
\begin{picture}(80,80)
\put(65,10){\makebox(0,0){$Q^2$,GeV$^2$}}
\put(20,77){\makebox(0,0){ $\eta$}}
\put(0,5){
\epsfxsize=8cm
\epsfysize=8cm
\epsfbox{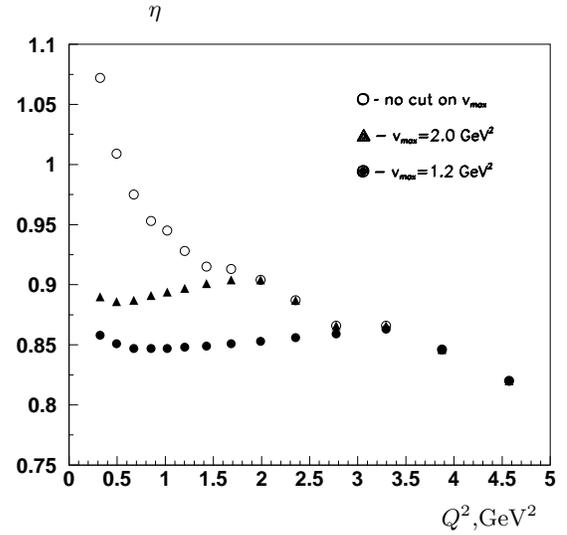}
}
\end{picture}
\vspace{-.9cm}
\caption{RC factor in kinematical conditions of HERMES
($e p \rightarrow e p \rho$).}
\label{f4}
}
\end{figure}

The dependence of $\eta$ on the kinematical variables $Q^2$, $W^2$ and
$t$ in the region of the collider experiments at HERA is presented in
Figure \ref{f3}. No cut on inelasticity is used, so $Q^2$-dependence
is similar to one given for the EMC/NMC experiment. It is found
that the RC factor $\eta$ is not sensitive for $W^2$. It is due to
the fact that
the photoproduction cross section is almost flat in the
kinematical region of the collider experiments.

The dependence on the $v_{m}$ cut in the region of HERMES kinematics is
analyzed
in Figure \ref{f4}. The usage of the cut changes RC factor  for small
$Q^2$ and does not influence on the RC for  larger values of $Q^2$.
It can be understood from Figure \ref{vmax}. In the case of the cut usage
we
have to define $v_m$ as a minimum of the value of the cut and $v_m$ given
by the kinematical restrictions (\ref{vmkin}).
For fixed values of $-t$ the cut influence on $v_m$ and RC till
certain value of $Q^2$ only.

\begin{figure}[t]
\unitlength 1mm
\begin{picture}(80,80)
\put(69,6){\makebox(0,0){$-t$,GeV$^2$}}
\put(20,70){\makebox(0,0){\small $\sigma, mb$}}
\put(5,0){
\epsfxsize=8cm
\epsfysize=8cm
\epsfbox{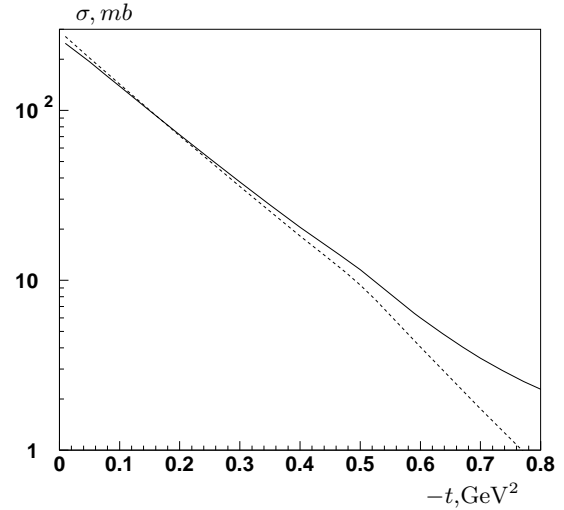}
}
\end{picture}
\vspace{-0.1cm}
\caption{The Born (dashed) and observed (solid) cross sections
in respect to $-t$. No cut on inelasticity is used.}
\label{tdep}
\end{figure}

\subsection{$t-$distribution}
\label{Num2}

\begin{figure}[t]
\unitlength 1mm
\begin{picture}(80,80)
\put(69,6){\makebox(0,0){$-t$,GeV$^2$}}
\put(20,70){\makebox(0,0){\small $b_v$}}
\put(5,0){
\epsfxsize=8cm
\epsfysize=8cm
\epsfbox{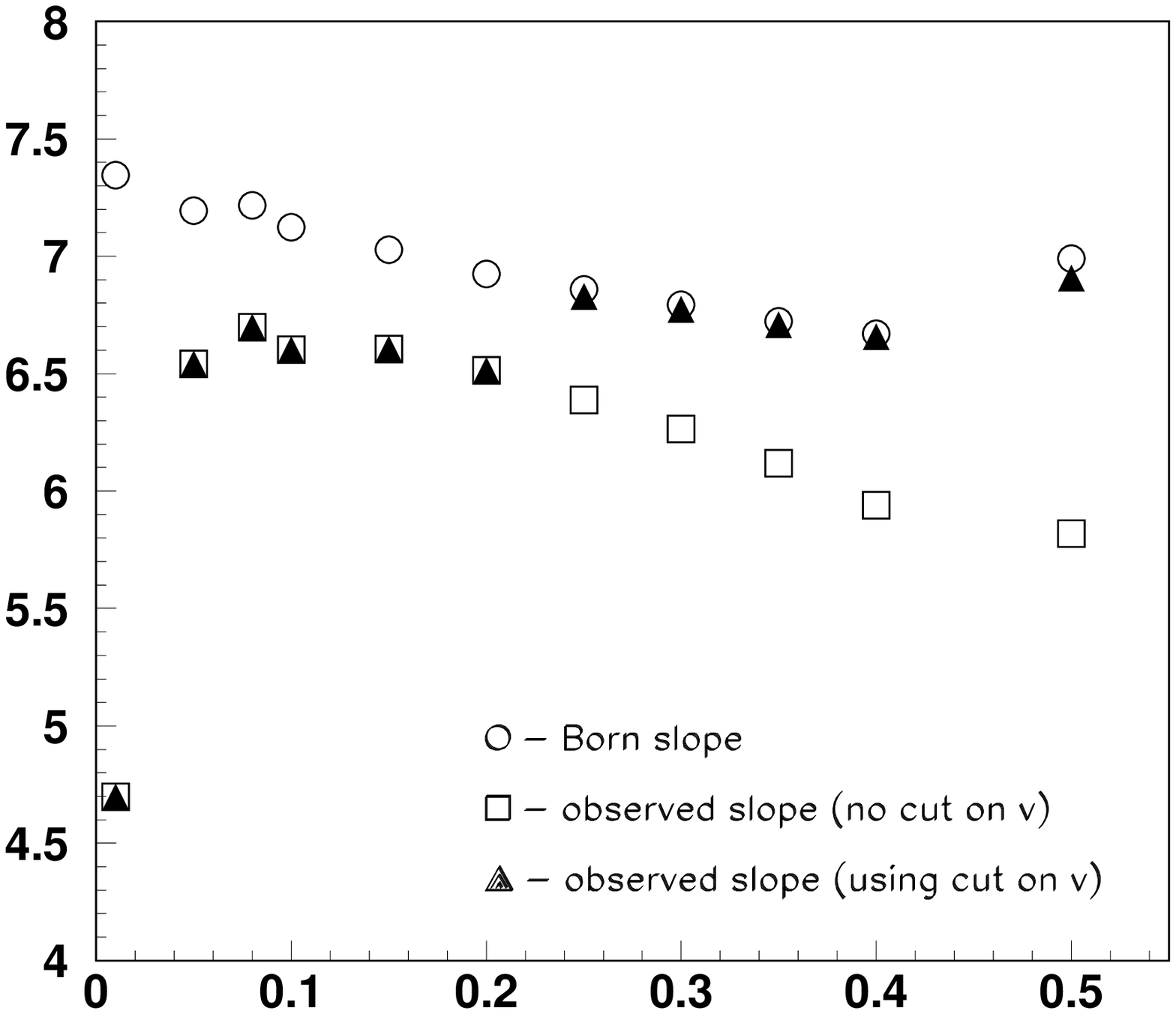}
}
\end{picture}
\caption{Slope parameter
in respect to $-t$}
\label{slope}
\end{figure}

The radiative effects can
influence on the slope of the observed cross section in respect to $-t$
due to essential dependence of RC on
this variable. It is illustrated in Figures \ref{tdep} and \ref{slope}.
The t-dependence of the Born cross section within the model of
refs.
\cite{Ryskin,AR} is
basically defined by two-gluon formfactor taken in
exponential
form $exp(b_vt)$ with $b_v=5$. The dependence on $p_t$ also influences  on
the slope
of the Born cross section in respect to $-t$. An additional dependence on
$t$ in the
case of RC comes basically from $v_m$, that is proportional to
$t-t_{min}$ in the diffractive region of small $-t$. Due to this fact
both
exponent $\exp(\delta_{inf})$ and the cross section $\sigma_F$
 tend to zero when $t \rightarrow t_{min}$. So the observed
cross section vanishes in this limit.

\subsection{Inelasticity distribution}
\label{Num3}

\begin{figure}[t]
\unitlength 1mm
\begin{picture}(80,80)
\put(69,6){\makebox(0,0){$v$,GeV$^2$}}
\put(20,70){\makebox(0,0){$N$}}
\put(5,0){
\epsfxsize=8cm
\epsfysize=8cm
\epsfbox{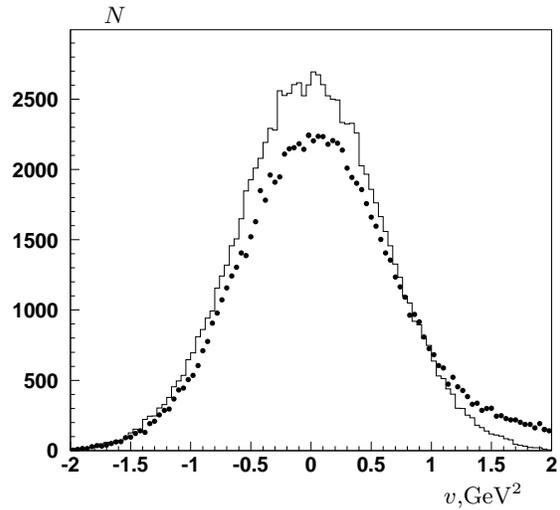}
}
\end{picture}
\caption{Inelasticity distributions: Born (line) and radiatively
corrected (circles)}
\label{f6}
\end{figure}

There are several reasons for the inelasticity to be non-zero: the finite
experimental
resolution, the  mixture of non-exclusive events and radiative effects.
In
this paper we assume that a special procedure of separation of
exclusive
and non-exclusive events has been performed. Strictly speaking
such
procedure can  suppress radiative effects partially. It depends on
the
details of data processing of a concrete experiment, and we assume
for simplicity that there are no such suppression.

Figure \ref{f6} gives two distribution: Born (pure Gaussian with a zeroth
mean value coming from the effects of a finite experimental resolution)
and
with taking into account RC. The second distribution consists of events of
two
types. First ones come from the Born, loop and radiated events with
$v<\sigma$. This distribution have the same mean value ($v=0$) and
$\sigma$ as the Born distribution. Other part of the events in
the radiative
corrected
distribution comes from the radiated events with $v>\sigma$. Total number
of
two distributions is the same.

\section{Discussion and Conclusion} \label{Dis}

In this article the QED radiative effects have been analyzed in kinematics
of recent experiments on the exclusive vector meson electroproduction.
The explicit covariant formulae for RC to the cross section are given in
eqs.(\ref{deltas}-\ref{020}). An approximate expression for
the bremsst\-rahlung
cross section can be found in eq.(\ref{051}).

The RC to the cross section of the diffractive vector meson
electroproduction is very sensitive to the cut on the inelasticity. Using
harder
cut leads to  smaller values of RC factor.

 In the diffractive region  ($ -t<0.3$) RC is negative
and
can reach
\begin{itemize}
\item {10\%} for muon experiments with fixed target;
\item {20\%} for electron experiments with fixed target;
\item {40\%} for electron collider experiments.
\end{itemize}
This large effect comes basically from a double logarithmic contribution
in
$\delta_{inf}$ (\ref{deltas}). For example for collider kinematics
both its logarithms
  exceed 10, and $\delta_{inf}$ can reach
0.5.

There are no essential dependence on the type of a vector meson. All
numerical results are given for the case of $\rho$ meson production.
The dependence on the type of a scattered lepton is typical.
RC in the case of the electron scattering is
several times more
due to the appearing the lepton mass
in the argument of the leading logarithm.

The observed cross section has steeper slope in respect to $t$
than
the Born cross section. RC to the slope parameter is
negative
and $\sim\penalty 10000 10\%$.

 FORTRAN code {DIFFRAD} is available ({aku@hep.by}) for the
calculation of RC to observable quantities in experiments on
the diffractive vector meson
electroproduction.

\section*{Acknowledgements}
I am grateful to A.Brull and N.Shumeiko for help and support. Also
I would like to thank N.Akopov, A.Borissov, A.Droutskoi, P.Kuzhir,
A.Nagaitsev, M.Ryskin, A.Soroko
and
H.Spiesberger for
fruitful discussions and comments.

\end{document}